\documentclass[conference]{IEEEtran}%TVT
\IEEEoverridecommandlockouts

\newif\ifarxiv\arxivfalse

\usepackage{cite}

\ifCLASSINFOpdf
\else
\fi

\usepackage{amsthm}

\usepackage{amsmath,amssymb,mathtools,dsfont,bm}

\usepackage{color}

\usepackage{algorithm,algorithmic}

\DeclarePairedDelimiter{\norm}{\lVert}{\rVert}

\usepackage{subcaption}

\makeatletter
\renewenvironment{thebibliography}[1]{%
	\@xp\section\@xp*\@xp{\refname}%
	\normalfont\footnotesize\labelsep .5em\relax
	\renewcommand\theenumiv{\arabic{enumiv}}\let\p@enumiv\@empty
	\vspace*{-2pt}% NEW
	\list{\@biblabel{\theenumiv}}{\settowidth\labelwidth{\@biblabel{#1}}%
		\leftmargin\labelwidth \advance\leftmargin\labelsep
		\usecounter{enumiv}}%
	\sloppy \clubpenalty\@M \widowpenalty\clubpenalty
	\sfcode`\.=\@m
}{%
	\def\@noitemerr{\@latex@warning{Empty `thebibliography' environment}}%
	\endlist
}
\makeatother

\let\OLDthebibliography\thebibliography
\renewcommand\thebibliography[1]{
	\OLDthebibliography{#1}
	\setlength{\parskip}{-0.1pt}
	\setlength{\itemsep}{-0.1pt}
}
\usepackage{xcolor}

\title{Robust Communication and Computation using Deep Learning via Joint Uncertainty Injection}

\author{\IEEEauthorblockN{Robert-Jeron Reifert\IEEEauthorrefmark{1}, Hayssam Dahrouj\IEEEauthorrefmark{2}, Alaa Alameer Ahmad\IEEEauthorrefmark{3}, Haris Gacanin\IEEEauthorrefmark{4} and Aydin Sezgin\IEEEauthorrefmark{1}}
	\IEEEauthorblockA{\IEEEauthorrefmark{1}Ruhr University Bochum, Germany, \hspace*{.2cm}\IEEEauthorrefmark{2}University of Sharjah, United Arab Emirates,}
	\IEEEauthorblockA{\IEEEauthorrefmark{3}Cariad SE, Wolfsburg, Germany, \hspace*{.2cm}\IEEEauthorrefmark{4}RWTH Aachen University, Germany.}
	\thanks{This work has been submitted to the IEEE for possible publication. Copyright may be transferred without notice, after which this version may no longer be accessible.\newline
		This work was supported in part by the German Federal Ministry of Education and Research (BMBF) in the course of the 6GEM Research Hub under grant 16KISK037.}
}

\begin{document}
	\maketitle
	
	\begin{abstract}
		The convergence of communication and computation, along with the integration of machine learning and artificial intelligence, stand as key empowering pillars for the sixth-generation of communication systems (6G). This paper considers a network of one base station serving a number of devices simultaneously using spatial multiplexing. The paper then presents an innovative deep learning-based approach to simultaneously manage the transmit and computing powers, alongside computation allocation, amidst uncertainties in both channel and computing states information. More specifically, the paper aims at proposing a robust solution that minimizes the worst-case delay across the served devices subject to computation and power constraints. The paper uses a deep neural network (DNN)-based solution that maps estimated channels and computation requirements to optimized resource allocations. During training, uncertainty samples are injected after the DNN output to jointly account for both communication and computation estimation errors. The DNN is then trained via backpropagation using the robust utility, thus implicitly learning the uncertainty distributions. Our results validate the enhanced robust delay performance of the joint uncertainty injection versus the classical DNN approach, especially in high channel and computational uncertainty regimes.\looseness-1
	\end{abstract}
	%\begin{IEEEkeywords}
	%Extended reality, cooperative NOMA, central cloud, cloud computing, mobile-edge computing, hybrid networks, uncrewed aerial vehicles.
	%\end{IEEEkeywords}
	
	% \fi
	\IEEEpeerreviewmaketitle

	\section{Introduction}
	The global mobile network data traffic is projected to surpass $550$ exabytes per month over the next few years \cite{emr}. This surge will involve substantial amounts of remote data connectivity and processing, thereby highlighting the need for an integrated communication and computation platform. Concurrently, the rise of delay-sensitive applications such as extended reality (XR) and digital twins is bound to further spearhead the need for advanced resource management techniques toward the development of the sixth-generation of communication systems (6G) \cite{9711524}. Notably, the International Telecommunication Union (ITU) envisions future 6G systems to natively incorporate machine learning (ML) and artificial intelligence (AI) as part of their design, deployment, and operation \cite{IMT-2030}. With ubiquitous intelligence, 6G would specifically be able to support the convergence of communication and computing, which is essential for XR video processing, as well as autonomous driving \cite{9061001}. Deep learning techniques, particularly, promise to deliver 6G networks solutions with low complexity, robustness, and near-optimality experience \cite{9061001}, e.g., provisioned wireless resource management \cite{10167900}, and robust optimization \cite{10071987}. Along such lines, this paper proposes, and evaluates the benefit of, using a deep neural network (DNN) to generate robust solutions for a complex joint communication and computation delay minimization problem.
	
	The problem addressed is related to the joint communication and computation resource management framework which is considered in the recent literature, e.g., \cite{10034763,10328648,6775376,6036193,10007803,9328305,9615112}. To this end, reference \cite{10034763} adopts a rate-splitting multiple access scheme and reference \cite{10328648} focuses on XR-empowered systems, yet, both \cite{10034763} and \cite{10328648} assume perfect knowledge of channels and computing requirements, which is challenging to realize in practice. From a robust resource allocation perspective, addressing the absence of channel state information has been considered in several works, e.g., \cite{6775376,6036193}. From a robust computation perspective, reference \cite{10007803} offers a framework to guarantee task deadlines under processing uncertainty. While references \cite{6775376,6036193,10007803} address only one type of uncertainty, i.e., either communication uncertainty or computing uncertainty, reference \cite{9328305} extends the scope to a joint perspective on computational intelligence in 6G. Further, \cite{9615112} proposes a robust edge computing application, where video streams are recorded and processed at the network edge before being streamed to the devices.
	
	Despite their numerical merits, \cite{10034763,10328648,6775376,6036193,10007803,9328305,9615112} adopt classical numerical optimization approaches. The synergy between 6G and AI/ML, however, nowadays provides powerful alternatives that promise to enhance seamless network performance, particularly in complex tasks related to advanced radio resource management problems \cite{10167900,10255239}. For instance, the optimization of communication and computation resources through ML-based algorithms is explored in \cite{huang2024joint} using deep reinforcement learning, and in \cite{9475121} using federated edge learning. As the available training data is never perfect, \cite{10071987} proposes using an uncertainty injection method that employs a deep learning-based architecture to maximize the network minimum-rate subject to imperfect channel state information. The perspective of jointly robust communication and computation, especially the one that captures their respective uncertainties, remains, however, largely unexplored in the literature, and so this paper addresses such issues using a joint uncertainty injection scheme.\looseness-1
	
	The paper proposes a robust deep learning framework to address the joint communication and computation resource management problem in the presence of channel and computing uncertainty. We consider a setup of one base station (BS) serving a number of devices using spatial multiplexing, i.e., via beamforming. Our focus is on minimizing the worst-case delay across all served devices by optimizing the transmit and computing powers, as well as the computing resources. %The formulated problem further considers interdependent communication and computing constraints, which stem from the channel and computing uncertainties. 
	Our proposed DNN-based approach takes estimated channels and computing requirements as inputs and produces optimized transmit and computing powers, along with computations allocations. The robustness of our solution is enhanced through a joint uncertainty injection framework, where the DNN's output undergoes further processing by injecting channel and computing error samples directly. 
	%	Under such a framework, the DNN parameters are updated by leveraging the $\gamma$-th quantile value, which represents the value above which a certain fraction ($\gamma$) of the worst-case delays under uncertainty injection occur. This quantile is utilized in the backpropagation process.
	Under such a framework, the DNN parameters are updated by leveraging the $\gamma$-th quantile value of worst-case delays under uncertainty injection utilizing backpropagation. 
	Our results highlight the enhanced robust delay performance of the joint uncertainty injection, especially for high parameters uncertainties and power-limited regimes. Explicit comparisons with two different uncertainty injection schemes and a state-of-the-art DNN-based solution show how the proposed algorithm is highly flexible, generalizes the uncertainty injection schemes, and provides enhanced robust delays.\looseness-1
	
	\section{System Model and Problem Formulation}\label{sec:sysmod}
	\begin{figure}[!t]
		\centering
		\includegraphics[width=\linewidth]{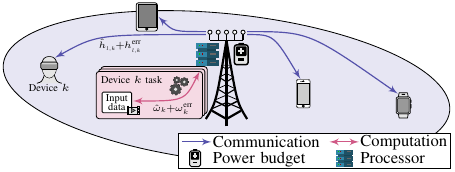}\\%3.2in
		\vspace*{-.2cm}
		\caption{Network setup with multi antenna BS and $4$ devices.}
		\label{fig:sys_mdl}
		\vspace*{-.5cm}
	\end{figure}
	%\textcolor{red}{Network Setup}\\
	In this work, we consider a single BS with $L$ antennas and embedded\hspace*{-.06cm} computing\hspace*{-.06cm} ability\hspace*{-.06cm} serving\hspace*{-.06cm} $K$\hspace*{-.06cm} single-antenna\hspace*{-.06cm} devices. We denote the antenna set by $\mathcal{L}=\{1,\cdots,L\}$, and the device set by $\mathcal{K}=\{1,\cdots,K\}$. An example of the considered model is illustrated in Fig.~\ref{fig:sys_mdl}. In fact, in many 6G-key applications, e.g., image and video processing or digital twinning, devices depend on strong remote processing capabilities at the network edge \cite{9328305}. %For instance, \cite{9615112} proposes \emph{UAVideo}, a framework where video data is generated at an uncrewed aerial vehicle, processed, and then forwarded to the devices.
	In such cases, each device experiences a delay consisting of two parts, namely the communication delay, i.e., the time loss on the physical communication medium, and the computation delay, i.e., the time for computing the device task.
	
	%\textcolor{red}{Channel Setup}\\
	\subsection{Channel Setup and Communication Delay}
	Let $h_{l,k}\in\mathbb{C}$ be the channel coefficient from the BS's antenna $l$ to device $k$, where $\bm{h}_{k} = [h_{1,k},\cdots,h_{L,k}]^T$ is the aggregate channel vector of device $k$. We assume that the BS has only access to an estimate of the channels. That is, the estimated channel from antenna $l$ to device $k$ is denoted by $\hat{h}_{l,k}\in\mathbb{C}$, with $k$'s aggregate estimated channel vector denoted by $\bm{\hat{h}}_{k} = [\hat{h}_{1,k},\cdots,\hat{h}_{L,k}]^T$. %The estimated channels approximate the real channels only up to some extend, with some uncertainty remaining. The related literature provides a multitude of different uncertainty models, examples include additive or multiplicative uncertainties \cite{6169188}, assuming specific uncertainty bounds \cite{6036193}, or opting for statistical uncertainties \cite{10071987}.
	In practice, the channel uncertainty distribution is not known \cite{6036193,9362239}. Yet, as channel estimations can be sampled in large amounts, a data-driven approach is able to \emph{learn} the channel uncertainty implicitly. Without loss of generality we model the channel coefficients as\looseness-1\vspace*{-.15cm}
	%	\phantom{bla}\vspace*{-.4cm}
	\begin{equation}
		h_{l,k} = \hat{h}_{l,k} + h^\text{err}_{l,k},\label{eq:hlk}%\vspace*{-.2cm}
	\end{equation}
	where $h^\text{err}_{l,k}\sim\mathcal{CN}(0,\sigma_h^2)$ is the channel estimation error.
	
%	We denote the digital precoder of device $k$'s signal at antenna $l$ by $v_{l,k}\in\mathbb{C}$. The beamforming vector of device $k$'s signal $\bm{v}_{k}\in\mathbb{C}^{L}$ is then defined by $\bm{v}_{k} = [v_{1,k},\cdots,v_{L,k}]^T$, where $\norm[]{\bm{v}_k}_2 = 1$.
	The beamforming vector of device $k$'s signal $\bm{v}_{k}\in\mathbb{C}^{L}$ is defined by $\bm{v}_{k} = [v_{1,k},\cdots,v_{L,k}]^T$, where %$v_{l,k}\in\mathbb{C}$ denotes $k$'s digital precoder at antenna $l$, and 
	$\norm[]{\bm{v}_k}_2 = 1$,
	adopting the regularized zero-forcing (RZF) strategy \cite{10071987}. We also let $p_k^\text{tx}$ be the BS's transmit power allocated to serve device $k$. The transmit power vector is then denoted by $\bm{p}^\text{tx} = [p_1,\cdots,p_K]$.
%	Similar to \cite{10071987}, we adopt the regularized zero-forcing (RZF) beamforming strategy. We also let $p_k^\text{tx}$ be the BS's transmit power allocated to serve device $k$. The transmit power vector is then denoted by $\bm{p}^\text{tx} = [p_1,\cdots,p_K]$.
	
	The devices messages are encoded into $\xi_k$, $\forall k\in\mathcal{K}$, where $\xi_k\sim\mathcal{CN}(0,1)$ are independent and identically distributed, circularly symmetric Gaussian random variables. Each device $k$ receives its own desired signal, interference from other signals, and noise. The received signal at device $k$ becomes\vspace*{-.1cm}
	\begin{align}
		y_k = \bm{h}_k^H \bm{v}_k \sqrt{p^\text{tx}_k} \xi_k + \sideset{}{_{j\in\mathcal{K}\backslash\{k\}}}{\sum} \bm{h}_k^H \bm{v}_j \sqrt{p^\text{tx}_j} \xi_j + n_k, \label{eq:yk}\\[-.65cm]\nonumber
	\end{align}
	where $n_k$ is the additive white Gaussian noise, with zero mean and variance $\sigma^2$. Based on \eqref{eq:yk}, the achievable rate of device $k$ is given by\vspace*{-.1cm}
	\begin{align}
		r_k = W\log_2\left(1+\frac{|\bm{h}_k^H \bm{v}_k|^2 p^\text{tx}_k}{\sum_{j\in\mathcal{K}\backslash\{k\}} |\bm{h}_k^H \bm{v}_j|^2 p^\text{tx}_j + \sigma^2}\right),\label{eq:rk}\\[-.6cm]\nonumber
	\end{align}
	where $W$ is the system bandwidth. Device $k$'s communication delay is then calculated as ${D_k^\text{out}}/{r_k}$, where $D_k^\text{out}$ is the output data size of the $k$-th device task computation. According to \eqref{eq:rk}, $r_k$ is a function of the channel values, which highlights the impact of the channel uncertainty on the communication delay.\looseness-1%\vspace*{-.2cm}
	
	\subsection{Computing Framework and Computation Delay}%\vspace*{-.1cm}
	From the computation perspective, device $k$ requests to process a task, which consists of an input data size $D_k^\text{in}$, and a computation intensity $\omega_k$, i.e., the required computation cycles per bit of input data is denoted by $\omega_k$, for a total of $\omega_kD^{in}_k$ computation cycles for device $k$'s task. However, such a computational requirement $\omega_k$ may not be perfectly known at the BS, as only an estimation $\hat{\omega}_k$ is available.
	Specifically, computing environments are highly dynamic, i.e., they vary depending on system load and concurrent processes, and systems dependency, i.e., the inter-relation of input data characteristics.
	Various statistical models for $\omega_k$ exist in the literature, including uniform, normal, and Gamma distributions \cite{lorch2001improving}.
	This paper addresses the computation uncertainty using a data-driven fashion, where we sample large amounts of uncertainty realizations and train the DNN accordingly. The computation intensity is modeled as\vspace*{-.1cm}
	\begin{equation}
		\omega_{k} = \hat{\omega}_{k} + \omega^\text{err}_{k},\label{eq:omegak}\vspace*{-.1cm}
	\end{equation}
	where $\omega^\text{err}_{k}\sim\mathcal{N}(0,\sigma_\omega^2)$ is the computing estimation error \cite{lorch2001improving}.
	
	With an estimation of the required computation cycles for device $k$'s task, i.e., $\hat{\omega}_k D_k^\text{in}$, the BS allocates $f^\text{co}_k$ in cycles/s to compute $k$'s task. The overall vector of computation allocations is denoted by $\bm{f}^\text{co} = [f^\text{co}_{1},\cdots,f^\text{co}_{K}]^T$. Due to the finite computation resources at the BS, the maximum computation capacity constraint is given by\vspace*{-.15cm}
	\begin{equation}
		\sideset{}{_{k\in\mathcal{K}}}\sum f^\text{co}_{k} \leq F^{\text{max}}, \label{eq:maxcompcap}\vspace*{-.15cm}
	\end{equation}
	where $F^\text{max}$ is the BS's maximum computation capacity. Due to the intertwined nature of communication and computation resources, especially in terms of the allocated power levels, we consider $p^\text{co}$ as the total power allocated by the BS for the computation of all tasks. The relation between the computation power $p^\text{co}$ and computation allocation $f_k$ is given by\vspace*{-.15cm}
	\begin{equation}
		p^\text{co} = \tau \left(\sideset{}{_{k\in\mathcal{K}}}\sum f^\text{co}_{k} \right)^{\mu} \label{eq:pco},\vspace*{-.1cm}
	\end{equation}
	where $\tau$ and $\mu$ are constants of the CPU model \cite{10328648}.
	
	By further considering the communication delay discussed earlier, device $k$ experiences a total delay $t_k$ consisting of a computation and a communication delay as follows\vspace*{-.15cm}
	\begin{align}
		t_k = \frac{\omega_k D_k^\text{in}}{f^\text{co}_k} + \frac{D_k^\text{out}}{r_k}. \label{eq:tk}\\[-.75cm]\nonumber
	\end{align}
	%	\begin{align}
		%		t_k = {\omega_k D_k^\text{in}}\left({f^\text{co}_k}\right)^{-1} + {D_k^\text{out}}\left({r_k}\right)^{-1}. \label{eq:tk}\\[-.75cm]\nonumber
		%	\end{align}
	In order to achieve fair service among all network devices, the paper aims at minimizing the worst-case delay $t_\text{max}$ among all devices, i.e., $t_\text{max} = \text{max}_{k\in\mathcal{K}}\{t_k\}$, computed via the ground-truth channels $\bm{h}_k$ and computation intensities $\omega_{k}$, $\forall k\in\mathcal{K}$.\vspace*{-.1cm}
	
	\subsection{Joint Communication and Computation Power}\vspace*{-.1cm}
	In this work, we consider the coupling of communication and computation power consumption (i.e., $\bm{p}^\text{tx}$ and $p^\text{co}$, respectively) at the BS, which leads to the overall power constraint\vspace*{-.15cm}
	\begin{align}
		\sideset{}{_{k\in\mathcal{K}}}\sum p_k^\text{tx} + p^\text{co} \leq P^\text{max}, \label{eq:P}\\[-.7cm]\nonumber
	\end{align}
	with total power budged $P^\text{max}$. Equation \eqref{eq:P} emphasizes the intertwined communication and computation powers, especially underlining the need for a careful selection of joint resource allocation under limited resources.\vspace*{-.2cm}
	
	%	The relationship between computing power $p^\text{co}$ and computing allocation $\bm{f}^\text{co}$ is depicted in equation \eqref{eq:pco}, while equation \eqref{eq:P} establishes the connection between computing power $p^\text{co}$ and transmit power $\bm{p}^\text{tx}$. While the channel and computing uncertainties directly influence the transmit power and computation allocation, respectively, they also indirectly impact each other. For instance, increased channel uncertainty may cause the algorithm to reserve more resources for communication, subsequently reducing the available computation power. This underscores the necessity for a unified protocol that manages the allocation of communication and computation resources, as proposed in this paper.
	
	\subsection{Robust Delay Utility}\vspace*{-.1cm}
	Based on \eqref{eq:tk}, one can infer that the overall delay $t_k$ of device $k$ is a function of the channel vectors $\bm{h}_k$ and the computation intensities $\omega_k$. Both $\bm{h}_k$ and $\omega_k$ follow some unknown probability distribution, such that the probability distributions of $t_k$ and $t_\text{max}$ are also unknown. Specifically, the BS only has access to the estimated $\bm{\hat{h}}_k$ and $\hat{\omega}_k$, $\forall k\in\mathcal{K}$.
	
	%	In classical optimization frameworks for wireless communication resource management \cite{10328648,10034763}, the BS treats $\bm{h}_k=\hat{\bm{h}}_k$, $\omega_{k}=\hat{\omega}_{k}$ and allocates the resources accordingly. In robust resource management, errors are often bounded and the worst-case channel or computing requirement is taken into account \cite{6036193}. In this work, however, we utilize
	To best account for the channel and computing uncertainty, the paper utilizes a DNN with inputs $\bm{\hat{h}}_k$, $\hat{\omega}_k$, $\forall k\in\mathcal{K}$, and inject samples of uncertainty realizations of both channels and computing requirements into the training phase just after the DNN output \cite{10071987}. That is, for a given resource allocation, we inject estimation errors to impact the objective value, and afterwards train the network according to the new (robust) objective.
	Such uncertainty injection supplies a multitude of potential worst-case delays $t_\text{max}$ for different realizations. This paper aims to minimize the $\gamma$-th quantile of the worst-case delay distribution, denoted by $t_\text{max}^\gamma$, where\vspace*{-.2cm}
	\begin{align}
		\text{Pr}\left[t_\text{max} > t_\text{max}^\gamma \big| \bm{\hat{h}}_k, \hat{\omega}_k, \forall k\in\mathcal{K} \right] \leq \gamma,\label{eq:t_percentile}\\[-.75cm]\nonumber
	\end{align}
	where $t_\text{max}^\gamma$ can be interpreted as the value for which only a fraction $\gamma$ of all realizations have higher worst-case delays.\vspace*{-.15cm}
	
	\subsection{Problem Formulation}\vspace*{-.1cm}
	The paper now aims at minimizing the robust worst-case delay $t_\text{max}^\gamma$ among all devices. The corresponding optimization problem is formulated as follows:
	
	\phantom{bla}\vspace*{-.7cm}
	\begin{subequations}\label{eq:Opt1}
		\begingroup
		\addtolength{\jot}{-.1cm}
		\begin{align}
			\underset{\bm{p}^\text{tx},p^\text{co},\bm{f}^\text{co}}{\text{min}}\quad &t_\text{max}^\gamma  \tag{\ref{eq:Opt1}} \\
			\text{s.t.} \quad\quad & \eqref{eq:hlk},\eqref{eq:omegak},\eqref{eq:maxcompcap},\eqref{eq:P}, \eqref{eq:t_percentile},\hspace*{-.2cm} \nonumber\\
			&t_\text{max} \geq t_k, &\forall k\in\mathcal{K},\label{eq:tmaxconstraint}\\[-.65cm]\nonumber
		\end{align}
		\endgroup
	\end{subequations}
	%	\begin{subequations}\label{eq:Opt1_}
		%			%\begingroup
		%			%\addtolength{\jot}{-.08cm}
		%			\begin{align}
			%					\underset{\bm{p}^\text{tx},p^\text{co},\bm{f}^\text{co}}{\text{min}}\quad &t_\text{max}^\gamma  \tag{\ref{eq:Opt1}} \\
			%					\text{s.t.} \quad\;\;\;\, %& \eqref{eq:P},\eqref{eq:maxcompcap},\hspace*{-.2cm} \nonumber\\
			%					&\text{Pr}\left[t_\text{max} > t_\text{max}^\gamma \big| \bm{\hat{h}}_k, \hat{\omega}_k, \forall k\in\mathcal{K} \right] \leq \gamma,\hspace*{-2.0cm} \label{eq:gammaconstraint}\\
			%					&h_{l,k} = \hat{h}_{l,k} + h^\text{err}_{l,k}, &\forall (l,k)\in(\mathcal{L},\mathcal{K}),\\
			%					&\omega_{k} = \hat{\omega}_{k} + \omega^\text{err}_{k}, &\forall k\in\mathcal{K},\\
			%					&t_\text{max} \geq t_k, &\forall k\in\mathcal{K},\label{eq:tmaxconstraint}\\
			%					&\sideset{}{_{k\in\mathcal{K}}}\sum f^\text{co}_{k} \leq F^{\text{max}}, \label{eq:compcapconstraint}\\
			%					&\sideset{}{_{k\in\mathcal{K}}}\sum p_k^\text{tx} + p^\text{co} \leq P^\text{max},\hspace*{-3cm} \label{eq:powerconstraint}
			%				\end{align}
		%			%\endgroup
		%		\end{subequations}
	where the optimization runs over the variables $\bm{p}^\text{tx}$, $p^\text{co}$, and $\bm{f}^\text{co}$, the computation capacity constraint is given in \eqref{eq:maxcompcap}, and the power constraint is given in \eqref{eq:P}. Constraint \eqref{eq:t_percentile} ensures the solution's robustness in terms of the $\gamma$-th quantile. Constraint \eqref{eq:tmaxconstraint} expresses the worst delay among all devices as $t^\text{max}$.
	The intricacy of problem \eqref{eq:Opt1} stems from the mathematical coupling between the communication and computation variables and constraints, as well as from the unknown distribution of the worst-case delay $t_\text{max}$ given the channel and computing states information uncertainties. The paper next proposes a data-driven solution approach based on a deep learning framework with joint uncertainty injection.\vspace*{-.1cm}
	
	%	\section{Proposed DNN-based Solution}\label{sec:dnn}
	\section{Deep Learning via Joint Uncertainty Injection}\label{sec:dnn}\vspace*{-.1cm}
	To tackle the intricate optimization problem \eqref{eq:Opt1}, we harness the promising abilities of deep learning in terms of automatic learning, feature representation, and versatility.
	More specifically, we first fix the beamforming vectors $v_k$ using: $\bm{V}=\bm{\hat{H}}(\bm{\hat{H}}\bm{\hat{H}}^H+\alpha\bm{I})^{-1}$, where $\bm{V}=[\bm{v}_1,\cdots,\bm{v}_K]$, $\bm{\hat{H}}=[\bm{\hat{h}}_1,\cdots,\bm{\hat{h}}_K]$, $\alpha$ is the RZF factor, and $I$ is a $K\times K$ identity matrix. Each $v_k$ is normalized to unit norm.
	%, i.e., $\bm{V}=(\bm{H}\bm{H}^H+\alpha\bm{I})^{-1}\bm{H}$ \textcolor{red}{multiple different (?) definitions}. %https://arxiv.org/pdf/2301.11515.pdf : W(HH^H+zI)^-1 H | https://ieeexplore.ieee.org/stamp/stamp.jsp?tp=&arnumber=1391204 : H*(HH*+aI)^-1 | WeiYu : H(H^H H+aI)^-1
	We then utilize a DNN to directly map problem \eqref{eq:Opt1}'s parameters to a resource allocation solution, which is then further processed by injecting uncertainties for optimizing a robust objective.
	In a typical DNN, neurons are organized into layers, including input, hidden, and output layers. These layers are densely interconnected, with each connection between neurons having an associated weight that is continuously adjusted during the training process. Additionally, activation functions are applied to the neurons within these layers, facilitating nonlinear transformations of their inputs. In general, depth (number of layers) and width (number of neurons in each layer) of the DNN are key factors that influence its capacity to learn and represent data.\looseness-1
	
	In the context of this paper, we propose a DNN architecture, which takes the estimated channels and estimated computing requirements as inputs, i.e., $\bm{\hat{h}}_k$ and $\hat{\omega}_k$, $\forall k\in\mathcal{K}$. Thus, the input layer consists of $2LK+K$ neurons, which represent the $2LK$ real and imaginary channel coefficients and the $K$ distinct devices' tasks. After the input layer, we attach $\mathcal{F}_\text{depth}$ fully-connected hidden layers with $\mathcal{F}_\text{width}$ neurons each, and Rectified Linear Unit (ReLU) activation functions. The output layer consists of $2K+1$ neurons, where the first $K+1$ entries represent the $K$ devices' allocated power and the BS's computation power, and the remaining $K$ entries represent the computation capacities. The DNN then returns the output vector $\bm{x}=[x_1,\cdots,x_{2K+1}]^T$, representing the optimized resource allocation $\bm{p}^\text{tx}$, $p^\text{co}$, and $\bm{f}^\text{co}$, after some mathematical manipulations.
	We let $\mathcal{F}_{\boldsymbol{\Theta}}$ denote the proposed DNN, such that problem \eqref{eq:Opt1} is solved by\vspace*{-.2cm}
	\begin{align}
		\bm{x} = \mathcal{F}_{\boldsymbol{\Theta}}(\bm{\hat{h}},\bm{\hat{\omega}}), \label{eq:dnn}\\[-.7cm]\nonumber
	\end{align}
	where $\bm{\hat{h}}=[\bm{\hat{h}}_{1},\cdots,\bm{\hat{h}}_{K}]^T$ and $\boldsymbol{\hat{\omega}}=[\hat{\omega}_{1},\cdots,\hat{\omega}_{K}]^T$. Note that $\boldsymbol{\Theta}$ refers to the DNN model parameters, e.g., weights, biases, etc.\footnote{In equation \eqref{eq:dnn}, the output $\bm{x}$ is a function of the DNN parameters $\boldsymbol{\Theta}$. Such DNN $\mathcal{F}_{\boldsymbol{\Theta}}$ and its parameters $\boldsymbol{\Theta}$ are sequentially updated via backpropagation as discussed in Section \ref{ssec:backprop} later.}
	\begin{figure}[!t]
		\centering
		\includegraphics[width=\linewidth]{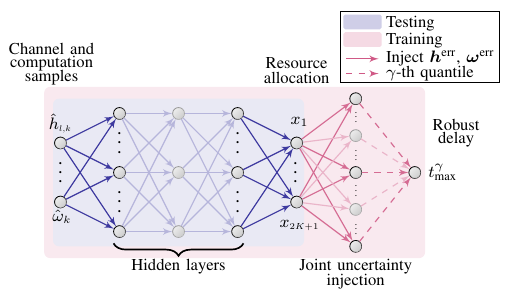}\\%3.2in
		\vspace*{-.4cm}
		\caption{Proposed DNN architecture with estimated parameters input, resource allocation output, and uncertainty injection.}
		\label{fig:dnn_mdl}
		\vspace*{-.5cm}
	\end{figure}
	Fig.~\ref{fig:dnn_mdl} illustrates a representation of the proposed DNN-based solution. %, \textcolor{red}{especially distinguishing training (red) and testing (blue) stages, whereas the training involves uncertainty injection.}
	In the context of minimizing the robust worst-case delay among devices, we next explore how to map the DNN output to the resource allocation, how to ensure the robustness of the solution, i.e., in terms of the $\gamma$-th quantile, and how to train the DNN according to the robust objective.
	
	\subsection{DNN Output to Resource Allocation Mapping}
	A major complication of the joint communication and computation resource management problem \eqref{eq:Opt1} is the interdependent computation capacity \eqref{eq:maxcompcap} and power \eqref{eq:P} constraints.
	We, thus, first use the softmax function\footnote{The softmax function of an input vector $\bm{x}$ is an output vector, the $i$-th entry of which is $e^{x_i}/(\sum_{j=1}^{n}e^{x_j})$, where $n$ is the dimension of $\bm{x}$.} on parts of the DNN output to compute the transmit and computation powers as\vspace*{-.1cm}
	\begin{equation}
		[(\bm{p}^\text{tx})^T,p^\text{co}]^T = P^\text{max}\cdot \text{softmax}\big([x_1,\cdots,x_{K+1}]^T\big).\label{eq:ptx_pco}\vspace*{-.1cm}
	\end{equation}
	By using the softmax function, we obtain $\sum_{i=1}^{K+1}x_i=1$, which ensures the full utilization of the complete power budget at the BS. Next, we transform the computation power into cycles using \eqref{eq:pco}, so as to obtain the maximum number of available computation cycles limited by the power allocation, denoted by $F^\text{pow}$. Based on \eqref{eq:pco}, $F^\text{pow}$ can be written as:\vspace*{-.1cm}
	\begin{equation}
		F^\text{pow} = \sqrt[\leftroot{1}\uproot{1}\mu]{\frac{p^\text{co}}{\tau}}.\vspace*{-.1cm}
	\end{equation}
	At this point, there are two limiting factors for the computation allocation (i.e., $\bm{f}^\text{co}$), namely, the BS's maximum computation capacity $F^\text{max}$ in \eqref{eq:maxcompcap}, and the above power-limited available computation cycles $F^\text{pow}$. We then use the smallest among the two limiting factors, i.e.,  $\text{min}\big(F^\text{max},F^\text{pow}\big)$. To determine the computation allocation $\bm{f}^\text{co}$, we use the softmax function on the remaining part of the output layer as follows\vspace*{-.1cm}
	\begin{equation}
		\bm{f}^\text{co} = \text{min}\big(F^\text{max},F^\text{pow}\big)\cdot \text{softmax}\big([x_{K+2},\cdots,x_{2K+1}]^T\big),\label{eq:fco}\vspace*{-.05cm}
	\end{equation}
	which utilizes all available computation cycles, either limited by the capacity or by the power.\vspace*{-.1cm}
	
	\subsection{Joint Uncertainty Injection Scheme}\vspace*{-.1cm}
	Given the resource allocation $\bm{p}^\text{tx}$, $p^\text{co}$, and $\bm{f}^\text{co}$ found in \eqref{eq:ptx_pco} and \eqref{eq:fco}, one can readily compute a worst-case delay for \emph{one given} channel and computing estimation. However, problem \eqref{eq:Opt1} aims at providing a robust $t_\text{max}^\gamma$ accounting for the uncertainties $\bm{h}^\text{err}=[h_{1,1}^\text{err},\cdots,h_{L,K}^\text{err}]^T$ and $\boldsymbol{\omega}^\text{err}=[\omega_1^\text{err},\cdots,\omega_K^\text{err}]^T$, given in \eqref{eq:hlk} and \eqref{eq:omegak}, respectively. %Thus, we next explain the uncertainty injection approach, first proposed in \cite{10071987} in the context of minimum rate maximization via power control, and present the joint uncertainty injection scheme adopted in the context of the current paper.
	Thus, we next explain the uncertainty injection approach \cite{10071987} and present the joint uncertainty injection scheme adopted in this paper's context.
	
	The basic idea of uncertainty injection is to estimate the statistical objective numerically, using a large number of realizations or samples of the uncertainty. Thereafter, such realizations are \emph{injected} after the DNN output to compute a large number of objective values. These objectives are then used to obtain an empirical estimate of the robust objective. Such estimate is then used to derive gradients and update the DNN parameters via backpropagation \cite{10071987}.
	
	In the context of this paper, however, the problem is subject to both channel and computing state information uncertainties. The adopted objective is also a function of the worst-case delay, as presented in \eqref{eq:t_percentile}, \eqref{eq:Opt1}. Therefore, we sample $N$ realizations of the channel and computing uncertainties $\bm{h}^\text{err}_n$, $\boldsymbol{\omega}^\text{err}_n$, $n\in\{1,\cdots,N\}$. These samples are used to compute $N$ worst-case delays $t_{\text{max},n}$, $\forall n\in\{1,\cdots,N\}$. With these delay-values derived from the uncertainty samples, we sort all $N$ delays and select the $(\gamma\cdot N)$-th highest value, which forms an empirical estimate of the $\gamma$-th quantile worst-case delay $t_\text{max}^\gamma$, also referred to as $\hat{t}_\text{max}^{\hspace*{.02cm}\gamma}$.%\vspace*{-.2cm}
	
	\subsection{Unsupervised Learning via Backpropagation}\label{ssec:backprop}\vspace*{-.1cm}
	An integral part of the proposed solution is the DNN training procedure, especially in accounting for the uncertainties to ensure a robust solution. More specifically, by taking the $(\gamma\cdot N)$-th highest value (i.e., $\hat{t}_\text{max}^{\hspace*{.02cm}\gamma}$) of all computed worst-case delays $t_{\text{max},n}$, $\forall n\in\{1,\cdots,N\}$, gradients are derived from $\hat{t}_\text{max}^{\hspace*{.02cm}\gamma}$ to update the DNN parameters. To be more specific, we use the gradient $\partial\hat{t}_\text{max}^{\hspace*{.02cm}\gamma}/\partial\Theta$, which is computed using Pytorch \cite{paszke2019pytorch}. In that way, the DNN is trained in an unsupervised fashion, implicitly learning the distribution of channel and computing uncertainties. The numerical merit of our proposed method is discussed in the simulations section of the paper, as shown next.\looseness-1\vspace*{-.2cm} %We omit the technical derivations of such updates for space limitation, especially that many of such derivations mirror the ones derived in \cite{10071987} and references therein.\looseness-1
	
	\section{Simulation Results and Discussion}\label{sec:sim}%\vspace*{-.1cm}
	In the following, we numerically evaluate the performance of the proposed DNN-based solution with joint uncertainty injection. The network consists of one BS with $L=8$ antennas serving $K=6$ devices. Other parameters are summarized in Tab.~\ref{tb:simparam}.
	The noise is set to $-75$ dBm/Hz. We also assume Rayleigh fading $h_{l,k}\sim \mathcal{CN}(0,1)$, $\forall (l,k)\in(\mathcal{L},\mathcal{K})$, utilize minimum mean-square estimation for the estimated channels $\bm{\hat{h}}$, and set the beamforming vectors $\bm{v}_k$, $\forall k\in\mathcal{K}$, according to RZF \cite{10071987}.
	As the beamforming vectors are determined a priori, we slightly modify the DNN's input layer, and only input effective channels, i.e., $\hat{h}_{i,j}^\text{eff}=\bm{\hat{h}}_i^H \bm{v}_j$, $\forall i,j\in\mathcal{K}$, thereby reducing the input layer to $2K^2+K$ nodes.
	The ground-truth computation intensities are drawn according to a Gamma distribution, i.e., $\omega_{k}\sim \Gamma(2,200)$, $\forall k\in\mathcal{K}$ \cite{10007803}.
	As for the DNN parameters and training procedure, we use $\mathcal{F}_\text{depth} = 10$ fully-connected hidden layers with $\mathcal{F}_\text{width} = 400$ neurons each, ReLU activation, train $500$ epochs with $50$ minibatches, each consisting of $1000$ independent network realizations. Validation and test sizes are set to $2000$ each, where each realization experiences uncertainty injection to obtain the robust objectives.
	\begin{table}[t]
		\vspace*{.2cm}
		\renewcommand{\arraystretch}{1.1}
		\centering
		% Some packages, such as MDW tools, offer better commands for making tables
		% than the plain LaTeX2e tabular which is used here.
		\begin{tabular}{c c c}
			\hline
			\hspace*{-.1cm}General\hspace*{-.1cm} & Communication\hspace*{-.3cm} & Computation\\%\hline
			\hline
			%	$K=6$ devices & $L=8$ antennas & \\
			\hspace*{-.1cm}$P^\text{max}=38$ dBm\hspace*{-.1cm} & $W=1$ MHz\hspace*{-.3cm} & $F^\text{max}=4.6\cdot 10^9$ cycles/s \\
			\hspace*{-.1cm}$N=1000$ samples\hspace*{-.1cm} & $D_k^\text{out}=7.5\cdot 10^4$ bits\hspace*{-.3cm} & $D_k^\text{in}=5\cdot 10^4$ bits\\
			\hspace*{-.1cm}$\gamma=0.05$\hspace*{-.1cm} & $\alpha=0.2$ RZF factor\hspace*{-.3cm} & $\tau = 10^{-28}$, $\mu = 3$\\\hline
		\end{tabular}%\vspace*{-6.0cm}
		\vspace*{-.15cm}
		\caption{Simulation parameters (unless otherwise mentioned).}
		\label{tb:simparam}
		\vspace{-.65cm}
	\end{table}%
	
	To benchmark the proposed joint uncertainty injection (referred to as \emph{Joint UI}), we utilize two modified schemes, \emph{Comm UI}, a method which only accounts for communication channel uncertainty, and \emph{Comp UI}, a method which only accounts for uncertainty in the computing requirements. Further, we consider the \emph{Regular DNN}, a scheme that does not account for the uncertainties at all, and is trained according to the estimated inputs only.\vspace*{-.2cm}
	
	\subsection{Impacts of the Uncertainties}
\begin{figure}[t]
		\centering
		\includegraphics[width=.9\linewidth]{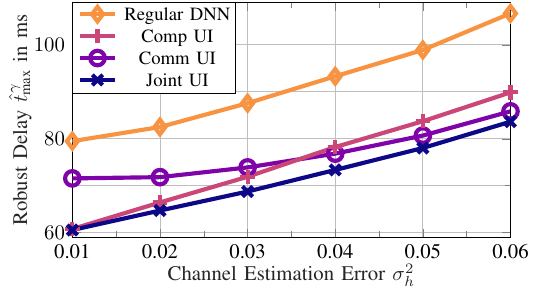}\\%3.2in
		\vspace*{-.3cm}
		\caption{Robust worst-case delay $\hat{t}_\text{max}^{\hspace*{.02cm}\gamma}$ as a function of the channel uncertainty, i.e., $\sigma_h^2$.}
		\label{fig:y_delay_x_commV}
		\vspace*{-.45cm}
	\end{figure}

\vspace*{-.1cm}
	In the first simulation set, we observe the impact of both channel and computing requirement uncertainties on the robust delay $\hat{t}_\text{max}^{\hspace*{.02cm}\gamma}$. To this end, Fig.~\ref{fig:y_delay_x_commV} depicts the robust delay versus the channel estimation error variance $\sigma_h^2$, while maintaining a fixed computation estimation error variance of $\sigma_\omega^2 = 6400$.
	Initially, it is noteworthy that all uncertainty injection schemes demonstrate improvements over the baseline \emph{Regular DNN}, with gains ranging between $8$ and $20$ ms.
	Notably, the \emph{Joint UI} scheme consistently achieves the lowest robust delay across all levels of channel uncertainty variances.
	Fig.~\ref{fig:y_delay_x_commV} highlights the generalizability and flexibility of \emph{Joint UI} in different uncertainty scenarios.
	Specifically, lower values of $\sigma_h^2$ mark a region where computing uncertainty dominates, while channel uncertainty has a lesser impact. Here, \emph{Comp UI} closely rivals \emph{Joint UI}, while \emph{Comm UI} performs relatively worse. In contrast, for higher channel uncertainties (e.g., $\sigma_h^2=0.06$), \emph{Comm UI} and \emph{Joint UI} exhibit similar performances, while \emph{Comp UI} shows a delay increase of over $6$ ms.
	A noteworthy comparison between \emph{Regular DNN} and \emph{Joint UI} can be drawn at $\hat{t}_\text{max}^{\hspace*{.02cm}\gamma}=80$ ms.
	\emph{Regular DNN} achieves $\hat{t}_\text{max}^{\hspace*{.02cm}\gamma}=80$ ms at approximately $\sigma_h^2=0.01$, whereas \emph{Joint UI} achieves lower delays, down to around $\sigma_h^2=0.053$. Consequently, when the robust delay is considered as a measure of delay guarantee, it becomes apparent that \emph{Joint UI} can guarantee better delays than \emph{Regular DNN}, even with significantly increased uncertainty variance. These findings underscore two significant insights: The importance of uncertainty injection in minimizing robust delays in general, and the effectiveness of the proposed \emph{Joint UI} scheme across various uncertainty scenarios in particular.
	
\begin{figure}[t]
		\centering
		\includegraphics[width=.9\linewidth]{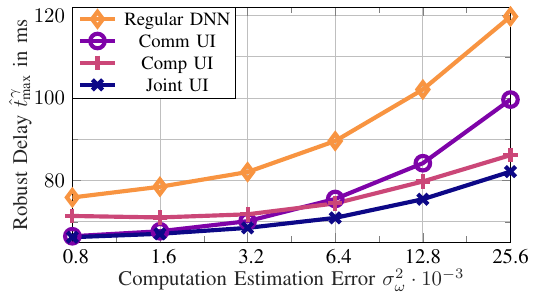}\\%3.2in
		\vspace*{-.3cm}
		\caption{Robust worst-case delay $\hat{t}_\text{max}^{\hspace*{.02cm}\gamma}$ as a function of computing uncertainty, i.e., $\sigma_\omega^2$.}
		\label{fig:y_delay_x_compV}
		\vspace*{-.55cm}
	\end{figure}

	% Next figure
	Fig.~\ref{fig:y_delay_x_compV} reaffirms the advantages of \emph{Joint UI} by showing the robust delay as a function of the computation estimation error variance $\sigma_\omega^2$, with $\sigma_h^2 = 0.035$. In Fig.~\ref{fig:y_delay_x_compV}, the high computation estimation error region ($\sigma_\omega^2>6400$) offers further insights. By comparing the slope of \emph{Regular DNN} and \emph{Comm UI} versus the slope of \emph{Comp UI} and \emph{Joint UI}, it is observed that both \emph{Comp UI} and \emph{Joint UI} exhibit a relatively small increase in performance degradation in response to high computing uncertainty. In fact, at $\sigma_\omega^2=25600$, the proposed \emph{Joint UI} achieves a gain of approximately $37$ ms over \emph{Regular DNN}, $17$ ms over \emph{Comm UI}, and $4$ ms over \emph{Comp UI}.
	
	All the examined uncertainty injection schemes in Fig.~\ref{fig:y_delay_x_commV} and Fig.~\ref{fig:y_delay_x_compV} notably improve the network robust delay performance, albeit with varying degrees of effectiveness for \emph{Comp UI} and \emph{Comm UI} across different uncertainty values. \emph{Joint UI}, particularly, showcases superior flexibility, generalizes both \emph{Comp UI} and \emph{Comm UI}, and achieves significantly reduced robust delays. These results highlight \emph{Joint UI}'s potential in spearheading the prospective joint communication and computation resource management schemes in future 6G networks.\vspace*{-.2cm}
	
	%	\begin{figure}[t]
		%		\centering
		%		\begin{subfigure}[t]{0.24\textwidth}
			%			\begin{center}
				%				\includegraphics[scale=0.66]{figures/res/y_delay_x_commV_v2}\vspace{-.3cm}
				%				\caption{Channel uncertainty.}%\vspace{-.1cm}
				%				\label{fig:y_delay_x_commV}
				%			\end{center}
			%		\end{subfigure}\hfill
		%		\begin{subfigure}[t]{0.24\textwidth}
			%			\centering
			%			\includegraphics[scale=0.66]{figures/res/y_delay_x_compV_v2}\vspace{-.3cm}
			%			\caption{Computing uncertainty.}%\vspace{-.2cm}
			%			\label{fig:y_delay_x_compV}
			%		\end{subfigure}\vspace*{-.2cm}
		%		\caption{Robust worst-case delay $\hat{t}_\text{max}^{\hspace*{.02cm}\gamma}$ as a function of the different uncertainties' variances, i.e., $\sigma_h^2$ and $\sigma_\omega^2$.} \label{bla2}\vspace*{-.6cm}
		%	\end{figure}
	
	\subsection{Joint Power Constraint}\vspace*{-.1cm}
	%	\begin{figure*}[t]
		%		\begin{subfigure}[t]{0.32\textwidth}
			%			\centering
			%			\includegraphics[width=.998\linewidth]{figures/res/y_delay_x_power_v2}%2.75
			%			\vspace*{-.3cm}
			%			\caption{Robust worst-case delay $\hat{t}_\text{max}^{\hspace*{.02cm}\gamma}$ vs. $P^\text{max}$.}\vspace{-.2cm} \label{fig:y_delay_x_power}
			%		\end{subfigure}\hspace{.1cm}
		%		\begin{subfigure}[t]{0.32\textwidth}
			%			\begin{center}
				%				\includegraphics[width=.96\linewidth]{figures/res/y_comm_comp_delay_x_power_v3}%2.75
				%				\vspace*{-.3cm}
				%				%				\caption{Communication (\emph{Comm}) and computation (\emph{Comp}) delay versus $P^\text{max}$.}
				%				\caption{Communication and computation delay.}
				%				\vspace{-.2cm} \label{fig:y_comm_comp_delay_x_power}
				%			\end{center}
			%		\end{subfigure}\hspace{.1cm}
		%		\begin{subfigure}[t]{0.32\textwidth}
			%			\centering
			%			\includegraphics[width=1.00\linewidth]{figures/res/y_delay_x_epochs.pdf}%2.75
			%			\vspace*{-.3cm}
			%			%\caption{\color{red}Min-rate over the number of devices comparing under- and overloaded networks.}\vspace{-.2cm}
			%			\caption{Convergence behavior.}\vspace{-.2cm}
			%			\label{fig:y_delay_x_epochs}
			%		\end{subfigure}%\hspace{.1cm}
		%		\caption{Robust worst-case delay $\hat{t}_\text{max}^{\hspace*{.02cm}\gamma}$ as a function of different system parameters comparing various baseline schemes.} \label{bla3}
		%		\vspace*{-.7cm}
		%	\end{figure*}
	\begin{figure}[t]
		\centering
		\includegraphics[width=.9\linewidth]{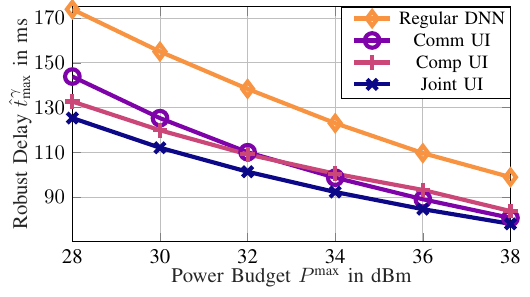}\\%3.2in
		\vspace*{-.25cm}
		\caption{Robust worst-case delay $\hat{t}_\text{max}^{\hspace*{.02cm}\gamma}$ vs. power budget $P^\text{max}$.}
		\label{fig:y_delay_x_power}
		\vspace*{-.45cm}
	\end{figure}
	\begin{figure}[t]
		\centering
		\includegraphics[width=.9\linewidth]{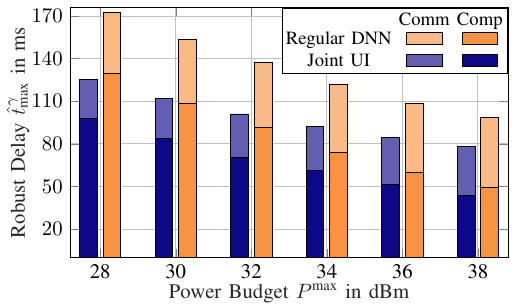}\\%3.2in
		\vspace*{-.2cm}
		\caption{Robust worst-case communication (\emph{Comm}) and computation (\emph{Comp}) delay vs. power budget $P^\text{max}$.}
		\label{fig:y_comm_comp_delay_x_power}
		\vspace*{-.55cm}
	\end{figure}
	%In practical systems, the total available power is an important design parameter. Specifically, according to \eqref{eq:P}, the power budget $P^\text{max}$ impacts both communication and computation variables. The following sets of simulations, therefore, capture such interplay and impact on the robust delay performance.
	%	In practical systems, the total available power plays a crucial role in design considerations. Specifically, according to \eqref{eq:P}, the power budget $P^\text{max}$ influences both communication and computation variables, thus affecting system performance. The following simulations capture this interplay and its impact on robust delay performance, with $\sigma_h^2=0.05$ and $\sigma_\omega^2=6400$.%\textcolor{red}{with $\sigma_h^2=0.025$ and $\sigma_\omega^2=3200$ / v2 $\sigma_h^2=0.05$ and $\sigma_\omega^2=6400$.}
	
	With $\sigma_h^2=0.05$ and $\sigma_\omega^2=6400$, Fig.~\ref{fig:y_delay_x_power} illustrates the robust delay $\hat{t}_\text{max}^{\hspace*{.02cm}\gamma}$ as a function of $P^\text{max}$.
	Across all schemes, $\hat{t}_\text{max}^{\hspace*{.02cm}\gamma}$ decreases with increasing power budget. With more power available at the BS, resources can be allocated more effectively, leading to improved delay performance.
	Notably, the proposed \emph{Joint UI}, \emph{Comm UI}, and \emph{Comp UI} schemes consistently outperform the \emph{Regular DNN} scheme, which neglects channel and computing uncertainties.
	For instance, in the low-power region ($P^\text{max}<32$ dBm), \emph{Joint UI} surpasses \emph{Regular DNN} by over $40$ ms, while this gain diminishes to around $15$ ms in the high-power region.
	Particularly in the low-power regime, \emph{Comm UI} exhibits significant performance degradation compared to \emph{Joint UI}, as it fails to consider computing uncertainty. In contrast, \emph{Comp UI} closely tracks \emph{Joint UI} with a nearly constant gap, highlighting the dominance of computation delay in low-power regimes.
	Once again, Fig.~\ref{fig:y_delay_x_power} emphasizes the necessity of accounting for both channel and computing uncertainties to optimize delay performance, particularly in the more constrained low-power regime.\looseness-1
	
	%	\begin{figure}[t]
		%		\centering
		%		%\includegraphics[width=.9\linewidth]{figures/res/y_comm_comp_delay_x_power_v2}\\%3.2in
		%		\includegraphics[width=.9\linewidth]{figures/res/y_comm_comp_delay_x_power_v3}\\%3.2in
		%		\vspace*{-.3cm}
		%		%\caption{Robust average (bars) and worst-case (lines) communication (\emph{Comm}, blue) and computation (\emph{Comp}, red) delay over the power budged $P^\text{max}$ for the proposed \emph{Joint UI} scheme.}
		%		%\caption{\emph{Joint UI}'s robust average and worst-case communication (\emph{Comm}) and computation (\emph{Comp}) delay versus $P^\text{max}$.}
		%		\caption{Robust worst-case communication (\emph{Comm}) and computation (\emph{Comp}) delay versus $P^\text{max}$.}
		%		\label{fig:y_comm_comp_delay_x_power}
		%		\vspace*{-.5cm}
		%	\end{figure}
	%Shifting our focus to the communication and computation delays of \emph{Joint UI}, Fig.~\ref{fig:y_comm_comp_delay_x_power} depicts both the average (among devices) and worst-case communication (${D_k^\text{out}}/{r_k}$) and computation (${\omega_k D_k^\text{in}}/{f^\text{co}_k}$) delays against the power budget $P^\text{max}$.
	To best illustrate the communication and computation delays of \emph{Joint UI} and \emph{Regular DNN}, Fig.~\ref{fig:y_comm_comp_delay_x_power} depicts both the worst-case communication (${D_k^\text{out}}/{r_k}$) and computation (${\omega_k D_k^\text{in}}/{f^\text{co}_k}$) delays against the power budget $P^\text{max}$.
	Firstly, from a computing perspective, Fig.~\ref{fig:y_comm_comp_delay_x_power} illustrates how an increasing power budget leads to reduced worst-case computation delays.
	On the other hand, the impact of $P^\text{max}$ on the communication delays is less pronounced.
	This discrepancy is reasonable, given that computation delays are, for this parameter choice, significantly higher than communication delays, thereby dominating the overall delay metric $\hat{t}_\text{max}^{\hspace*{.02cm}\gamma}$.
	The imbalance between communication and computation delays is particularly evident in the low-power region. For instance, for \emph{Joint UI} at $28$ dBm, the computation delay is $98$ ms, whereas the communication delay is only $27$ ms. However, as more power becomes available (as depicted on the right-hand side of Fig.~\ref{fig:y_comm_comp_delay_x_power}), the \emph{Joint UI} scheme allocates additional resources favoring computation delay reduction. Consequently, computation delays approach communication delays as $P^\text{max}$ increases.\looseness-1\vspace*{-.15cm}
	
	%	\subsection{Convergence Behavior}\vspace*{-.15cm}
	%	%	\subsection{Convergence and Cumulative Distribution}
	%	%	\begin{figure}[t]
		%		%		\centering
		%		%		\includegraphics[width=.9\linewidth]{figures/res/y_delay_x_epochs.pdf}\\%3.2in
		%		%		\vspace*{-.3cm}
		%		%		\caption{Convergence behavior of the considered schemes.}
		%		%		\label{fig:y_delay_x_epochs}
		%		%		\vspace*{-.5cm}
		%		%	\end{figure}
	%	Lastly, we show the considered schemes' convergence behavior in Fig.~\ref{fig:y_delay_x_epochs}, with $\sigma_h^2=0.035$ and $\sigma_\omega^2=6400$ for $1000$ epochs. All considered schemes are shown to converge within the observed epochs. The major difference in the convergence plots can be seen when comparing \emph{Joint UI} and \emph{Comm UI} versus \emph{Comp UI} and \emph{Regular DNN}. The former schemes, i.e., \emph{Joint UI} and \emph{Comm UI}, both account for the channel uncertainty during training. The resulting convergence behavior seems more \emph{noisy} for both training and validation performance. Additionally, both schemes encounter a steep decrease of the objective value between epoch $100$ and $200$. %, after which especially \emph{Joint UI} becomes almost constant.
	%	Such convergence \emph{jump} is less obvious for \emph{Comp UI} and \emph{Regular DNN} and appears much earlier.
	%	Such behavior leads to the conclusion that communication uncertainty is the more severe factor for this specific parameter set, which is validated by the results of Fig.~\ref{fig:y_delay_x_commV}, where \emph{Comm UI} outperforms \emph{Comp UI}.\looseness-1\vspace*{-.1cm}

	\section{Conclusion}\label{sec:con}%\vspace*{-.2cm}
	Automating the decision making via ML algorithms is among the most prominent key enablers of future seamless networks. This paper introduces a novel ML-aided approach to address the convergence of communication and computation in 6G systems, while taking uncertainties in channel estimation and computing requirements into consideration. Our proposed robust solution effectively manages transmit and computing powers, alongside computation allocation, using a DNN-based approach. By injecting uncertainty samples during training and optimizing resource allocations based on robust utility, our approach minimizes the worst-case delay among multiple devices. Our findings demonstrate the superior performance of joint uncertainty injection for various power budgets and uncertainty levels, thereby unveiling the proposed solution potential at promoting robust communication and computation resource management schemes in future 6G systems.%\vspace*{-.2cm}
	
	\bibliographystyle{IEEEtran}
	\bibliography{bibliography}
	
\end{document}